\documentclass{ijcaArticle}
\usepackage{amssymb} \usepackage{amsmath}
\usepackage{graphicx} 
\usepackage{algorithm} \usepackage{algorithmic}
\usepackage{graphics}
\usepackage{ulem}

\newtheorem{Proposition}{Proposition}
\usepackage{amssymb}
\usepackage{verbatim}
\usepackage{caption}

\ijcaVolume{62}
\ijcaNumber{14}
\ijcaYear{2013}
\ijcaMonth{January}

\ijcaVolume{62}
\ijcaNumber{14}
\ijcaYear{2013}
\ijcaMonth{January}

\begin{document}

\title{Schedulability Analysis of Distributed Real-Time applications under Dependence and Several Latency Constraints}

\author{ 
   \large Omar KERMIA \\[-3pt]
   \normalsize CDTA  \\[-3pt]
    \normalsize Algiers, Algeria \\[-3pt]
    \normalsize okermia@cdta.dz \\[-3pt]
   }

\terms{Distributed Systems, Real-Time Operatin Systems}

\maketitle

\begin{abstract}
This paper focuses on the analysis of  real-time non
 preemptive multiprocessor scheduling with
precedence and several latency constraints.
 It aims to specify a schedulability condition which 
 enables a designer to check a priori -without executing or simulating-
  if its scheduling of
tasks will hold the precedence between tasks as well as several latency constraints imposed on determined pairs of tasks. 
It is shown that the required analysis is closely linked to the topological structure of the application graph. More precisely, it depends on the
configuration of tasks paths subject to latency constraints.
As a result of the study, a sufficient schedulability condition is introduced for precedences and latency constraints in the hardest configuration in term of complexity with an optimal number of processors in term of applications parallelism. In addition, the proposed conditions provides a practical lower bounds for general cases.  
Performances results and  comparisons with an optimal approach demonstrate the effectiveness of the proposed approach.   
\end{abstract}

\section{Introduction}
\label{}
Nowadays, computer applications in which computation must satisfy stringent timing constraints are widespread. In such applications, failure to meet the specified
deadlines can lead to a serious degradation of the system, and can also result in catastrophic loss of life or property. The increasing of computing requirements leads to the distribution 
of real-time applications over multi-core platforms. However, in addition to the complexity of parallelizing such applications, system designers  are faced to the problem of how to deal with applications parameters in such a way that their temporal constraints are met. Yet, the formalization of the performance of parallelisable applications date to year 1967 with the Amdahl law \cite{Amdahl67} and which was followed by a large number of works one of them is in \cite{Shi1996}.      

The challenge is to ensure that the real-time requirements of distributed applications are satisfied by providing formal methods. In order to schedule, a scheduling
algorithm is required which includes a set of rules defining the execution of tasks at the system runtime. At the same time, it is important to provide a schedulability analysis, which determines, whether a
set of tasks with parameters describing their temporal behavior will meet their temporal constraints. The result of such a test is typically a yes or a no. This answer indicates whether,
the constraints will be satisfied or not. These schemes and tests demand precise assumptions about task properties, which hold for the entire system lifetime.  In addition,  a set of processors are available for executing a set of distributed real-time applications or software. Each computing element might be a processor in a multi-processor architecture, a host or a core in a multi-core machine. Without loss of generality, the term  \lq{}processor\rq{} is used in the present paper instead of the other ones.

In this paper, a theoretical study is performed for solving the problem of analyzing a system of real-time tasks under precedence and several latency constraints. Latency constraints
addressed in this work are that imposed by the system designer between predefined pairs among tasks of the application graph.   Latency constraints analysis can be used to test, both at design time and for on-line execution, whether the time lapses between tasks pairs executions does not exceed an already specified values and, so,  meet their deadlines. It constitutes a serious alternative to extensive testing and simulation by providing analytical latency bounds which contribute considerably in process monitoring and control applications required by real-time performance guarantees.

As it is mentioned previously, the paper is interested in non-preemptive scheduling. This choice is motivated by a variety of reasons including \cite{jeffaynonpreemptive}:

\begin{itemize}
\item \ In many practical real-time scheduling problems such as I/O scheduling,
properties of device hardware and software either make preemption impossible or
prohibitively expensive. The preemption cost is either not taken into account or
still not really controlled;

\item \ Non-preemptive scheduling algorithms are easier to implement than preemptive
algorithms, and can exhibit dramatically lower overhead at runtime;

\item \ The overhead of preemptive algorithms is more difficult to characterize and
 predict than that of non-preemptive algorithms. Since scheduling overhead is
 often ignored in scheduling models, an implementation of a
 non-preemptive scheduler will be closer to the formal model than an
 implementation of a preemptive scheduler.
\end{itemize}

For these reasons, designers often use non-preemptive approaches, even though
elegant theoretical results on preemptive approaches do not extend easily to them \cite{SERTS}.  Designers also choose directed acyclic graphs (DAG) to model different kinds of structures in mathematics and computer science. Indeed, in many real time systems, applications are developed using DAGs \cite{10.1109/RTCSA.2011.72} where vertices represent sequential code segments and edges represent precedence constraints. Throughout the paper, it is explained that the latency constraint is strongly linked to the topology of the applications graph or more accurately to the parts of the graph concerned by latency constraints.

There is a large literature in the real-time community on scheduling tasks on multi-processor architectures. Sporadic and aperiodic real-time tasks are considered in respectively  \cite{DBLP:conf/rtss/BaruahG08}  
and \cite{DBLP:conf/ecrts/BaruahL04} whereas energy-efficient scheduling is proposed in \cite{DBLP:conf/rtcsa/HuangCT11}. In \cite{journals/rts/AbeniCLMP05} QoS management is proposed and \cite{DBLP:journals/tii/ButtazzoBW11} targets to minimize either the overall
bandwidth consumption or the required number of cores. However, to our knowledge, schedulability analysis dealing with several latency constraints (as it is defined in this paper) has not been considered. In fact, Among the constraints addressed in real-time scheduling issues, latency constraints are less studied comparing with the periodicity constraint for example \cite{papierok}. Nevertheless, latency is a major concern in several fields such as in embedded signal processing applications \cite{Goddard98onthe} for example. In the literature, most often, authors talk about an end-to-end deadline which ensures that the time lapse from sensors and actuators does not exceed a certain value \cite{Hsueh_Lin_2000}. The main differences between latency and  end-to-end deadline is that latency constraints are as much as system designer wants meaning that they can be imposed between any pair of connected tasks in the system (not necessarily sensor and actuator tasks only). 
In \cite{aor07}, a definition of this constraint is given and the existence of a link between deadlines and latency is proven. In addition, distributed architectures involve inter-processor communications the cost of which must be taken into account accurately. Furthermore, concerning synchronization cost reduction, the approach proposed in \cite{Chao94minimizingredundant} is efficient in term of finding a minimal set of interprocessor synchronization, however, this approach
assumes that some dependence can be removed even though data are exchanged. Moreover, it is not suitable for latency constraints satisfaction because it imposes a tasks scheduling not exploiting the potential tasks parallelism which is essential in minimizing their total execution time.   Moreover, it was not possible to exploit results from parallelism community, essentially because of precedence constraints which are not taken into account \cite{Cucinotta11-SOMRES}.

The main contributions of this paper are
the proposition of a schedulability conditions for latency constraints in the hardest configuration with an optimal number of processors in terms of application parallelism. This configuration stands for the hardest configuration among the other possible configurations because of the interdependence of latency constraints. Also, from these conditions, practical lower bounds  for latency constraints values were deduced, the efficiency and the rapidity of which were showed by evaluation tests.

The paper is organized as follows: Section 2 introduces the model and defines the latency constraint. Section 3 introduces the schedulability analysis  through the different possible cases. Section 4 describes the performance evaluation.

\section{Definitions and Model}\label{model}
The paper deals with systems of real-time tasks with precedence and several latency constraints. A task $t_i$ is characterized by a worst case execution time (WCET)
$C(t_i) \in {\mathbb N}$. The precedences between tasks are represented by a directed acyclic graph (DAG) denoted $\cal G$ such that ${\cal G}=(\mathbb{V},\mathbb{E})$. $\mathbb{V}$ is the
set of tasks characterized as above, and $\mathbb{E} \subseteq \mathbb{V}\times \mathbb{V}$ the set of edges which represent the precedence (dependence) constraints between tasks. Therefore, the directed pair of tasks $(t_a,t_b)\in \mathbb{E}$ means that $t_b$ must be scheduled, only if $t_a$ was already scheduled and $t_a$ is called a predecessor of $t_b$.  The set of tasks belonging to all paths from $t_a$ to $t_b$ including $t_a$
and $t_b$ is  denoted by ${\mathbb V}\rq{}$.  Note that the architecture plate-form is composed of identical processors.
  
A communication cost is involved when dependent tasks are scheduled on two processors, whereas, the communication cost is considered to be negligible if dependent tasks are scheduled on the same processor.   In our study the overall communication overhead involved by the interaction between processors is taken into account.  If ${\cal M}$ is the function of time needed for communication then ${\cal M}$  can vary linearly with the number of processors:  ${\cal M}(m) = Q . (m-1)$ where $Q$ is a constant dependent on the architecture and stands for an average communication cost between a pair of processors and $m$ is the number of processors. In addition,   ${\cal M}$ can, also, vary logarithmically since communications can be designed in order to get a logarithmic impact on the total execution time. For example, communications can be parallelized in the case of hierarchical topology architectures and function ${\cal M}$ becomes ${\cal M}(m)  = Q.\log{m}$. Nevertheless, it is important to notice that in targeted applications, granularity is chosen in such a way to get high computation to communication ratio. 
Because, when the granularity is large the computation cost becomes dominant  and the relatively small (but non-negligible) communication cost actually encourages the use of more processors to help the reduction of scheduling time. This implies more opportunity for performance increase but, nevertheless, involves hard efficient load balancing \cite{Book:IntroductionToParallelComputing}.

Each task $t_i$ has a start time $S(t_i)$ determined by the scheduling algorithm.  A latency constraint is defined only between two tasks connected in the tasks graph which means that it exists at least one path connecting the two tasks. By imposing a latency constraint
$L({t_a,t_b})$, the time elapsing from the execution start of $t_a$ and the execution start of $t_b$ must be less or equal than an integer denoted also by $L({t_a,t_b})$ and which is already known. As in the graph tasks $t_a$ and $t_b$ are connected by one or several paths, hence, $\boldsymbol{{\cal P}(t_a,t_b)}$ denotes the set of paths $\boldsymbol{p_i}$ which connect $t_a$ to $t_b$.  Hence, ${\cal P}(t_a,t_b)$ is also a set of sets of tasks meaning that $t_i \in (p_j \in {\cal P}(t_a,t_b))$. 

The length  of $p_i$ is denoted by $\boldsymbol{|p_i|}$ such that $|p_i| = \sum_{t_j \in p_i} C(t_j)$. Among paths $p_i$,  $\boldsymbol{ lp}$  denotes the longest one.  

More formally, a latency constraint  $L({t_a,t_b})$ is met if and only if:
\begin{equation}
  S(t_b) - S(t_a) \leq L  
\end{equation}

\begin{figure}[h!] 
\begin{center} 
\includegraphics[scale=0.46]{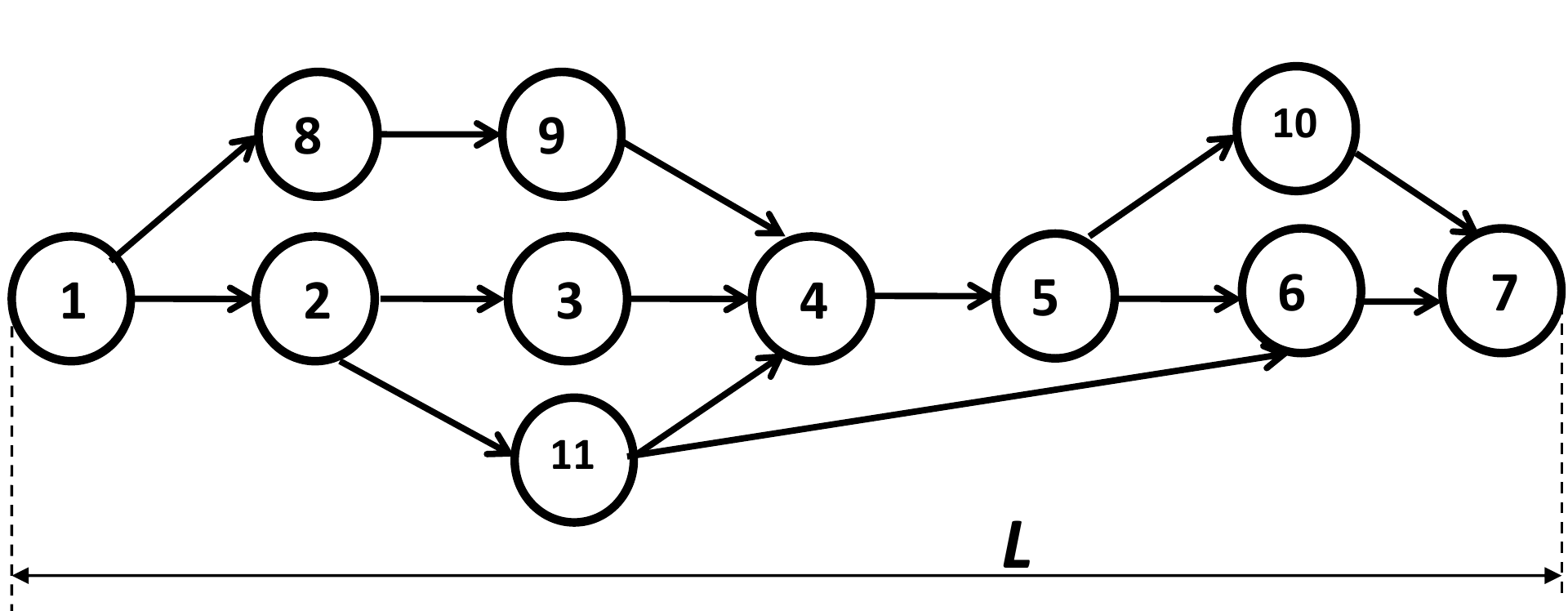} 
\caption{ Tasks under latency constraint} 
\label{cons-clust} \end{center} 
\end{figure}

In the tasks graph of the figure \ref{cons-clust}  ${\cal P}(t_1,t_7)$=$\{p_1, p_2, p_3, p_4, p_5, p_6\}$ such that: $ p_1$ = \{$t_1$, $t_2$, $t_3$, $t_4$, $t_5$, $t_6$, $t_7$\},  $p_2$= \{$t_1$, $t_8$, $t_9$, $t_4$, $t_5$, $t_6$, $t_7$\},  $p_3$=\{$t_1$, $t_2$, $t_3$, $t_4$, $t_5$, $t_{10}$, $t_7$\}, 
$p_4$= \{$t_1$, $t_8$, $t_9$, $t_4$, $t_5$, $t_{10}$, $t_7$\}, $p_5$=\{$t_1$, $t_2$, $t_{11}$, $t_4$, $t_5$, $t_6$, $t_7$\}, $p_6$=\{$t_1$, $t_2$, $t_{11}$, $t_4$, $t_5$, $t_{10}$, $t_7$\} and $p_7$=\{$t_1$, $t_2$, $t_{11}$, $t_6$, $t_7$\}.

\section{Schedulability study}
The studied problem is close to the problem "P $|$ prec $|$ $C_{max}$" (using Lenstra's 3-fields notation \cite{grah79}) which is known to be
NP-hard \cite{grah79}. The "P $|$ prec $|$ $C_{max}$" problem aims to minimize maximum completion time of all tasks whereas the objective is to determine the schedulability of the graph tasks by findingl whether a scheduling of all tasks of the graph on a multiprocessor platform, satisfying the precedence and latency constraints, exists or not. Consequently, our problem in a one latency case is also NP-hard. Moreover, in the several latency constraints case, the problem becomes NP-hard in the strong sens because of links between latency constraints .   

Since the studied problem is NP hard, no algorithm can resolve it in a polynomial time (unless NP=P) and this is, also, true for the schedulability condition. This means that, in a
general case, it is impossible to propose a necessary and sufficient condition allowing to check if a set of tasks under a latency constraint is schedulable or not in a polynomial time.

\subsection{One latency Constraint Case}

The matter of dealing with a latency
constraint is closely linked to the structure of the graph. That is the reason why a partitioning method is proposed considering graph paths. Without loss of generality, in the present paper it is considered that the whole graph is under the latency constraint $L(t_a,t_b)$ which means that the considered graph has one root vertice $t_a$ and one leaf vertice $t_b$ (see figure \ref{cons-clust}). In the case of graphs with large tasks and edges numbers, the number of paths is also very large. However, determining all paths is not an NP hard problem \cite{VisSchKonBor10}. Besides, according to \cite{Tutzauer2007}, it exists several approaches for determining all paths of a graph, among which the topological sort of the graph can be mentioned. However, in practice, the number of paths is less than the number of vertices in a graph. Even in a simple design with a small quantity of components, the number of vertices in $\cal G$ is more than 10 times the number of paths in the architecture \cite{YuchunMa:2007}.

The allocation algorithm (Algorithm \ref{algo1}) has as inputs all paths of the graph and as outputs the selection of some of them which, each one, will be associated to a distinct processor. First, the algorithm begins by sorting paths in ${\cal P}(t_a,t_b)$ according to a decreasing order of their lengths then it selects them one by one and it allocates paths tasks to a processor to which it is associated. After that, at each step, tasks belonging to a path $p_i$ and which were not allocated before via another path (the case of tasks belonging to several paths) will be allocated to the processor to which $p_i$ is associated. The algorithm stops when all tasks under a latency constraint are allocated meaning that all paths will not be necessarily selected.   

As a result, each task of the application graph will be allocated to only one processor. Also, an integer $m$ is returned equivalent to the number of selected paths which returns 
the number of required processors. In other words, Algorithm \ref{algo1} parallelizes the execution of the application by allocating its tasks to a set of processors. Besides, this parallelization follows the configuration of paths which compose the application graph.    



\begin{algorithm} 
\caption{Allocation Algorithm} 
\label{algo1} \begin{algorithmic}[1] 
\STATE  $m \leftarrow 0$ 
\STATE Sort paths in ${\cal P}$ in a decreasing order of length
\STATE Select $lp$  and initialize a set of tasks $\Phi = lp$  
\WHILE{$\Phi \neq \mathbb{V\rq{}}$} 
\STATE For each path $p_i$ not already selected : \\ $\lambda(p_i) = \sum\limits_{\substack{t_j \in p_i \ \wedge \ t_j \notin \Phi}} C(t_j)$ 
\STATE Select $p_i$ such that $\lambda(p_i) =\max (\lambda) $ 
\STATE $\Phi = \Phi \cup p_i$  \ \ \ \ \ \ (include $p_i$\rq{}s tasks in  $\Phi$)
\STATE $m \leftarrow m + 1$
\ENDWHILE 
\end{algorithmic} 
\end{algorithm}

\begin{figure}[h!] 
\begin{center} 
\includegraphics[scale=0.46]{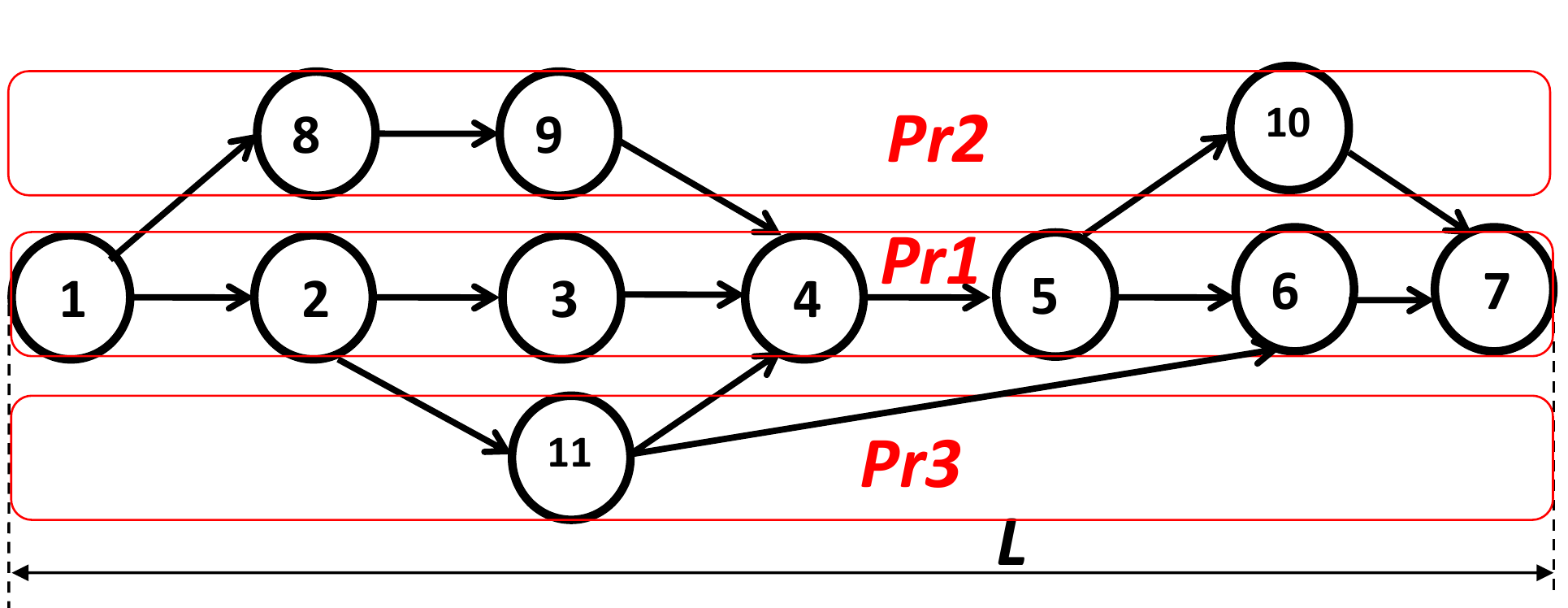} 
\caption{Paths Allocation} 
\label{cons-clust-exam} \end{center} 
\end{figure}
 
An example of Algorithm \ref{algo1} application is illustrated in figure \ref{cons-clust-exam}. Processors $P_1$, $P_2$ and $P_3$ were required whereas seven paths were detected (see example of section \ref{model}). For this example it is assumed that the execution times of tasks are equal.  From now the set of paths 
${\cal P}(t_a,t_b)$ is considered composed of $m$ paths (the ones selected by Algorithm \ref{algo1}). Also, we notice by $\widehat{p_i}$ the set of tasks exclusively belonging to $p_i$, more formally, if $t_i \in \widehat{p_i}$  then $\forall p_j \in {\cal P}(t_a,t_b) \setminus p_i , t_i \notin p_j$.

One can ask what makes the number of tasks returned by Algorithm \ref{algo1} so distinctive. The answer is that the value of $m$ represents the optimal number of processors since it allows to exploit the total parallelism inherent to the application graph. This means that if two tasks are not linked by a path in the graph (no one is the predecessor or the successor of the other) then
they are allocated to distinct processors. Moreover,  Adding other processors than the $m$ processors required by Algorithm \ref{algo1} does not improve the exploitation of the parallelism inherent to the application graph. Proposition \ref{gamma} introduces the optimality of $m$.

\begin{Proposition}\label{gamma} 
The application of Algorithm \ref{algo1} on an application graph returns the optimal number of processors allowing the task parallelism exploitation. 
\end{Proposition}
\vskip2mm 
\textbf{Proof}
Algorithm \ref{algo1} allocates tasks according to paths to which they belong.  Notice that the considered paths are those which include, at least, a task which does not
belong to any other path.
Let assume that for a given graph $G$ algorithm \ref{algo1} returned $m$ processors. Also, let assume that, it exists a number of processors $m \lq{}$ such that $m \lq{} < m$ for which the exploitation of the parallelism of the graph $G$ is optimal.  
This means that each pair of tasks $(t_{i},t_j)$ not linked by a path in $G$ are allocated to two distinct  processors among the $m \lq{}$ processors. As assessed earlier, the graph $G$ has only one root task $t_a$ and only one leaf task $t_b$ and, hence, it exist two distinct paths which link $t_a$ and $t_b$ and include $t_i$ for the first and $t_j$ for the second (This is due to the fact that $t_i$ and $t_j$ are not linked). This implies that all distinct paths in $G$ will be concerned.  Consequently, ($m - m \lq{}$) processors are missing in order to parallelize all pairs $(t_{i},t_j)$ $\boxdot$

From now on, Algorithm \ref{algo1} is systematically applied to allocate tasks. 
The following proposition introduces a necessary and sufficient schedulability condition in the case of one latency constraint.
\vskip2mm
\begin{Proposition}\label{clus_sim_n_cond_2} 
 Let $L$ be a latency constraint imposed on the tasks pair $(t_{a},t_{b})$. Latency constraint $L(t_{a}, t_{b})$  is met if and only if:  \ \ $\forall p_j \in {\cal P}(t_a,t_b), $
\begin{equation}\label{eq_clus_sim_cond_multi_2} 
\sum\limits_{t_i \in \ \bigcap p_j }^{} C(t_i) +  \max\limits_{{p_j}}(\sum\limits_{t_i \in \widehat{p_j}}^{} C(t_i)) + { \cal M}(m)  \leq L 
\end{equation} 
\end{Proposition}
\vskip2mm
\textbf{Proof} 
this result is quite intuitive and can be obtained by examining the inequality $ S(t_b) - S(t_a) \leq L$. Indeed $ S(t_b) - S(t_a)$ which is the scheduling time of tasks under latency constraint $L$ is equal to the sum of execution times of:
\begin{enumerate}
\item \ Tasks which are  non-parallelisable with any other tasks (sequential tasks which are linked by a path in the application graph). These are represented by tasks shared between all paths in ${\cal P}(t_a,t_b)$ ($t_i \in \ \{\bigcap p_j\} $), 
\item \ Among parallel tasks, the longest sub-path is selected from the $m$ paths. On each processor $m_i$ tasks of the set $\widehat{P_i}$  are allocated and the
largest sum of executions time of tasks of each  $\widehat{P_i}$ is kept. This is due to the precedence between tasks which prevents of distributing parallel tasks between processors to get a more balanced distribution such as $\frac{\sum\limits_{t_i \in {\cal V}\rq{}} C(t_i)}{m}$ (${\cal V}\rq{}$ is the set of tasks which are in parallel in the graph application), 
\item \ Communication overhead$\boxdot$ 
\end{enumerate}

\subsection{Several Latency Constraints Case}

In \cite{cucu04}, authors have stated that all possible combinations for two pairs of tasks under,
each one, a latency constraint can be covered by three cases:
\begin{itemize}
\item \ In parallel, when there is no path linking tasks under the first latency constraint to those under the second latency constraint.

\item  \ In $Z$, when there is one (or more) path linking tasks under the first latency to those under the second latency or vice versa. 
\item  \ In $X$, there is one (or more) path linking tasks under the first (resp. second) latency to those under the second (resp. first) latency. 
\end{itemize}
For the $Z$ and parallel relations the schedulability study can be performed as for the one latency case. This statement issues from the fact that latency constraints in these cases can be addressed one after the other in order to check the schedulability of the whole system. In addition, the $X$ configuration is the hardest one to be studied because the two latency constraints are dependent. In fact, satisfying one of these latencies is not related to the scheduling of tasks under this constraint only but it is related, also, to some tasks which are under other latency constraints. Usually, in this case, it is about multi-objective optimization and the problem becomes harder than in a single optimization case \cite{GlaBer:2010}.

\begin{figure}[h!] 
\begin{center} 
\includegraphics[scale=0.55]{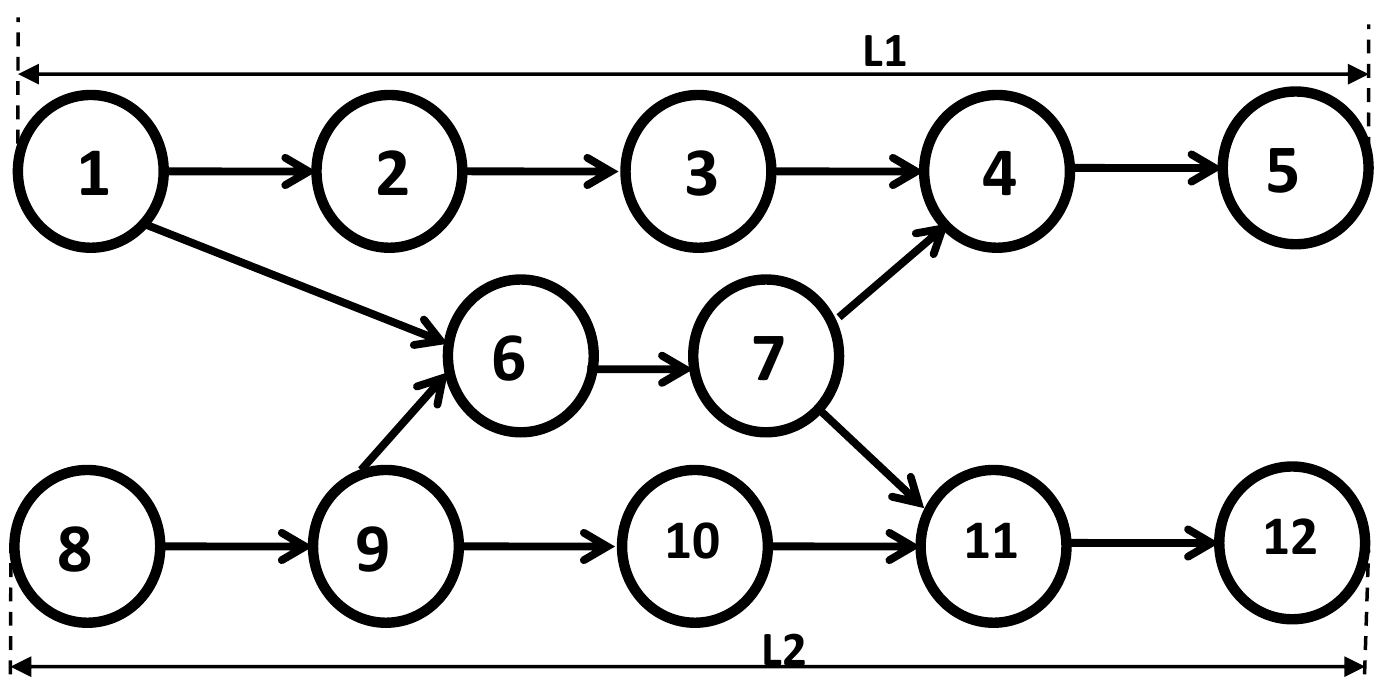} 
\caption{A pair of latency constraints in $X$} 
\label{fig-lat-crois} 
\end{center} 
\end{figure}

Let's take an example of a tasks graph subject to  a pair of latency constraints in $X$. The figure \ref{fig-lat-crois} depicts a pair of latency constraints $L_1$ and $L_2$ in $X$ imposed between $(t_1,t_4)$ and $(t_9,t_{11})$.

The following proposition introduces a necessary and sufficient schedulability condition in the case of two latency constraints in $X$.
\vskip2mm
\begin{Proposition}\label{th_X}
  Let $(L_1,L_2)$ be two latency constraints in $X$ imposed, respectively, on tasks pairs $(t_{a},t_{b})$ and $(t_{c},t_{d})$. Latency constraints $L_1(t_{a}, t_{b})$  and $L_2(t_{c}, t_{d})$ are met if and only if: 
\begin{enumerate}
\item Condition of proposition \ref{clus_sim_n_cond_2} is met for tasks under $L_1$ and $m_1$ processors and for tasks under $L_2$ and $m_2$ processors
\item  and

\begin{eqnarray}
\left \{\begin{array}{l}
          \max\limits_{p_i \in  {\cal P}(t_c,t_b)} |p_i| +  {\cal M}(m)   \leq L_1      \label{equ_x1} \\
   \max\limits_{p_i \in {\cal P}(t_a,t_d)} |p_i|  +   {\cal M}(m)  \leq L_2                    \label{equ_x2} 
\end{array} \right.
\end{eqnarray}
\end{enumerate}

\end{Proposition} 

$m$, $m_1$ and $m_2$ are obtained by applying Algorithm \ref{algo1} on the graph under latency  constraints $L_1$ and $L_2$. $m_1$ is the number of processors to which tasks under $L_1$ are allocated, $m_2$ the number of ones to which tasks under $L_2$ are allocated and $m$ represents all required processors. Notice that $m < m_1 + m_2$ because there exist tasks under the two latency constraints.


\vskip2mm

\textbf{Proof} 
As expected, the one latency case schedulability condition (condition \ref{eq_clus_sim_cond_multi_2}) becomes a necessary condition in the case of two latency constraints in $X$. Indeed, if one of the two latency constraints is not met then all the system is considered as non-schedulable.
Then, in order to prove the sufficiency of the condition proposed here, equations (\ref{equ_x1}) is assumed as satisfied, and constraints $L_1$ and $L_2$ are, nevertheless,  not met.  The constraints $L_1$ and $L_2$ are not met means that $S(t_b)-S(t_a) > L_1$ and $S(t_d)-S(t_c) > L_2$. 
\\\\
$S(t_b)-S(t_a) > L_1$ means that:

Either, 

$\exists p_i \in {\cal P}(t_a, t_b), \ |p_i| +  {\cal M}(m_1) > L_1$. This hypothesis is in contradiction with the condition \ref{eq_clus_sim_cond_multi_2}  because:

$ (\ref{eq_clus_sim_cond_multi_2}) \Rightarrow  \forall p_i \in {\cal P}(t_a,t_b), \  |p_i|  \leq L_1$     \\
\vskip1mm
Or, \\ as $t_c$ is a predecessor of the task $t_b$, hence, the start execution of $t_b$ is related to the execution of $t_c$ and other tasks which are under the latency constraint $L_2$. Therefore,  in the present case, the start execution of $t_b$ is delayed by the execution   
of tasks under latency constraint $L_2$ whereas all predecessor tasks of $t_b$ under latency constraint $L_1$ were executed. This is, more formally, described by the following inequality: \\ 
$\exists t_x \in ({\cal P}(t_a,t_b) \cap  {\cal P}(t_c,t_d))$, 
\begin{equation}\label{equ_x3}
 \max\limits_{\substack{p_i \in {\cal P}(t_c, t_x)}} |p_i| +  {\cal M}(m_2) <  \max\limits_{\substack{p_i \in {\cal P}(t_a, t_x)}} |p_i| +  {\cal M}(m_1) 
\end{equation}

Furthermore,\\ $S(t_b)-S(t_a) > L_1$ and (\ref{equ_x3})  $\Rightarrow$ 

$\  \exists  p_j \in {\cal P}(t_c,t_x) \ \ $ and$  \ \ \exists p_k \in {\cal P}(t_x,t_b), $
\begin{equation}\label{equ_x4}
|p_j|+ {\cal M}(m_1) + |p_k| + {\cal M}(m_2)  > L_1
\end{equation}

Otherwise, it is clear that:  
\begin{equation}\label{equ_x5}
|p_j|+ {\cal M}(m_1) + |p_k| + {\cal M}(m_2) \leq \max\limits_{p_i \in  {\cal P}(t_c,t_b)} |p_i| +  {\cal M}(m) 
\end{equation}
from condition \ref{equ_x1}, equation \ref{equ_x4} is in contradiction with equation \ref{equ_x5}.  

\vskip1mm

\noindent The same reasoning can be followed to prove that $(S(t_c)-S(t_d) > L_2)$ is in contradiction with the assumption that the
constraint $L_2$ is met$\boxdot$
\vskip2mm
The result of proposition \ref{th_X}is easily generalizable to a tasks graph subject to $n$ latency constraints, two by two, in $X$ configuration. Indeed, It suffices to
check  conditions of proposition \ref{th_X} for each pair of latency in $X$ then to conclude the schedulability of the whole system.  So, using results of propositions \ref{clus_sim_n_cond_2} and \ref{th_X} any application graph can be dealt with whatever the number of imposed latency constraints is and whatever these latency constraints are configured.

\vskip4mm
The schedulability study performed earlier introduces schedulabilty conditions over a processors number which stands for the optimal number to exploit all the parallelism inherent to the application graph, but the proposed conditions does not fit a system with a static architecture (i.e., the number of processors is known beforehand and fixed). When system designers face such systems, 
they tend towards fast analysis methods even thought these methods are not as exact as optimal methods. So, knowing that the targeted problem is NP-hard in the strong sens the schedulability analysis of such systems throughout optimal approaches or even heuristics takes a very long time.  Instead of an optimal schedulability analysis, conditions the paper proposes practical lower bounds for latency constraints values $L_i$ whatever the number of processors is. Hence, system designers can refer to the proposed conditions 
to adjust the latency constraints values while saving a considerable time.  The following proposition introduces lower bounds for latency constraints values according to the different configurations.

 \begin{Proposition}\label{lower_bound}
1.  if $L$ is a latency constraint imposed on the tasks pair $(t_{a},t_{b})$. The lower bound of $L(t_{a}, t_{b})$  is: 
\begin{equation}\label{eq_clus_sim_cond_multi_4} 
L^{l b}= \sum\limits_{t_i \in \ \bigcap p_j }^{} C(t_i) +  \max\limits_{{p_j}}(\sum\limits_{t_i \in \widehat{p_j}}^{} C(t_i)) + { \cal M}(m) 
\end{equation}

2. If $(L_1,L_2)$ are two latency constraints in $X$ imposed, respectively, on tasks pairs $(t_{a},t_{b})$ and $(t_{c},t_{d})$. The lower bounds of $L_1(t_{a}, t_{b})$  and $L_2(t_{c}, t_{d})$ are:

 \begin{eqnarray}\nonumber \label{lb}
 \begin{array}{l}        
L_{1}^{l b}= \boldsymbol{\max (}\sum\limits_{t_i \in \ \bigcap p_j }^{p_j \in {\cal P}(t_a,t_b)} C(t_i) +  \max\limits_{p_j \in {\cal P}(t_a,t_b)}^{} (\sum\limits_{t_i \in \widehat{p_j}}^{} C(t_i)) + { \cal M}(m_1) \boldsymbol{,} 
\\ \ \ \ \ \ \ \ \ \ \ \ \ \ \ \ \ \ \ \ \ \ \ \ \ \ \ \ \ \ \ \ \ \ \ \ \ \ \ \ \max\limits_{p_j \in  {\cal P}(t_c,t_b)} |p_j| +  {\cal M}(m)\boldsymbol{)}   \\   
\ \ \ \ \ \ \ \ \ \ \ \ \ \ \ \ \ \ \ \ \ \ \ \ \ \ \ \ \ \ \ \ \ \ \ \ \ \ \ \ \ \ \ \ \ \ \ \ \ \ \ \ \ \ \ \ \ \ \ \ \ \ \   \ \ \ \ \ \ \ \ \ \ \ \ \ \ \ \ \ \ \ \ \ \ \ \ \ \ \ \ \ \ \ \ \ \ \ \ \ \ \   (8)\\
 L_{2}^{l b}=  \boldsymbol{\max (}\sum\limits_{t_i \in \ \bigcap p_j }^{p_j \in {\cal P}(t_c,t_d)} C(t_i) +  \max\limits_{p_j\in {\cal P}(t_c,t_d)}^{}(\sum\limits_{t_i \in \widehat{p_j}}^{} C(t_i)) + { \cal M}(m_2)\boldsymbol{,} 
 \\   \ \ \ \ \ \ \ \ \ \ \ \ \ \ \ \ \ \ \ \ \ \ \ \ \ \ \ \ \ \ \ \ \ \ \ \ \ \ \ \  \max\limits_{p_j \in {\cal P}(t_a,t_d)} |p_j|  +   {\cal M}(m)\boldsymbol{)}  
 \end{array} 
\end{eqnarray}
\end{Proposition}
 
\textbf{Proof} \\
$L^{l b}$ represents a lower bound to the scheduling time between $t_a$ and $t_b$ ($S(t_b)-S(t_a)$). This means that the value that system designer will give to $L(t_{a}, t_{b})$ must not be lower than $L^{l b}$ otherwise the latency will necessarily be not met.  As high computation applications are targeted, the use of more processors involves the reduction of scheduling time. Reciprocally the reduction of the number of processors will increase the scheduling time. This proves that $l^{lb}$ in the different seen configurations is a minimum of scheduling time
for systems where the number of processors is less than $m$. 

In addition, as $m$ represents the optimal number of processors to get the optimal parallelism within the application graph, the fact of using more processors than $m$ processors
does not lead to reduce the scheduling time$\boxdot$

\subsection{Performance Evaluation}\label{first_tests}
 In order to evaluate the performances of applying the schedulability condition of proposition \ref{th_X} we implemented an application designated as the proposed approach which, for a given graph of tasks under a pair of latency constraints in $X$, checks conditions of proposition  \ref{th_X} and outputs, following the obtained result, the schedulability of the system.  Then, two kinds of tests are performed:
\begin{itemize}
\item \ an evaluation of time performances of the proposed solution, 
\item \ a comparison with solutions provided by the constraint programming approach.
\end{itemize}

Tasks graphs (DAGs) used for the evaluation were generated randomly according to the two following parameters: number of tasks  and density. In our case the graph density is a ratio between the number of edges in the graph and the number of  possible edges (in the complete graph). For example, a graph of 12 tasks with 0.5 density has 33 edges whereas
a complete graph of 12 tasks has 66 edges. Notice that the number of edges in a complete graph is $\frac{n(n-1)}{2}$, where $n$ is the number of tasks.       

Inside the graph, 40 \% of tasks is put under the constraint $L_1$ and 40 \%  under the constraint
$L_2$. Next, the remaining 20\%  are put under the two constraints $L_1$ and $L_2$.  An example of a generated graph with 12 tasks and 0.25 of density (17 edges ) is given in figure \ref{rand}: 5 tasks are exclusively under the constraint $L_1$, 5 other tasks are exclusively under $L_2$  and 2 tasks are under both of $L_1$ and
$L_2$. 

In the generated graph the number of edges is determined by the density (as explained in the previous paragraph) whereas the configuration of these edges is defined randomly as follows:

\begin{figure}[h!] 
\begin{center} 
\includegraphics[scale=0.58]{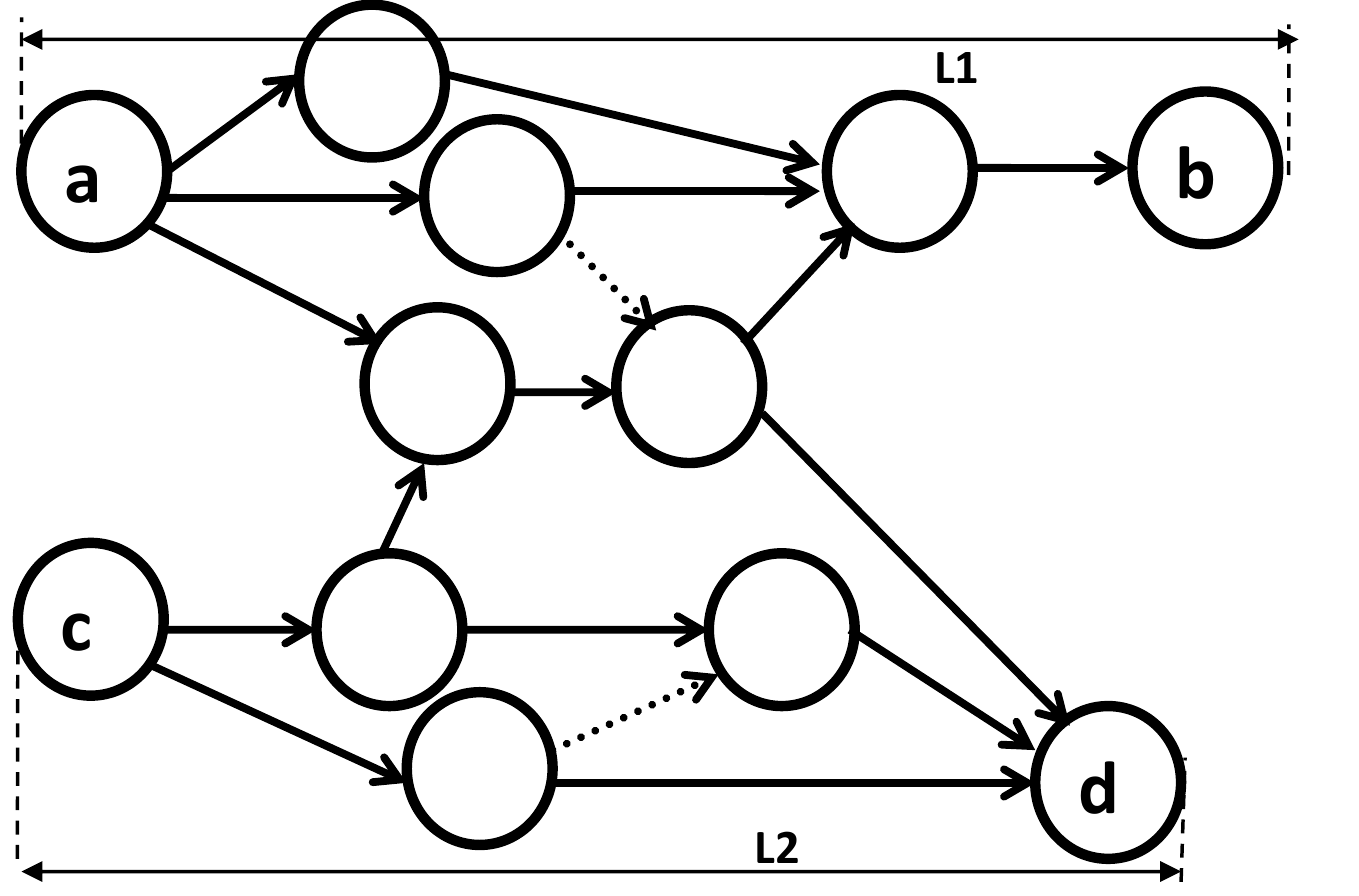} 
\caption{Example of generated 12 tasks graph} 
\label{rand} 
\end{center} 
\end{figure}

\begin{itemize}
\item \ a set of randomly generated edges within the restriction of ensuring the $X$ configuration of latency constraints (the edges in continued line in the graph of figure \ref{rand}),   
\item \ a set of randomly generated edges between tasks under the same latency constraints (the edges in discontinued line in the graph of figure \ref{rand}) and which
satisfy the DAG properties of the graph.
\end{itemize}

\begin{figure}[h!] 
\begin{center} 
\includegraphics[scale=0.70]{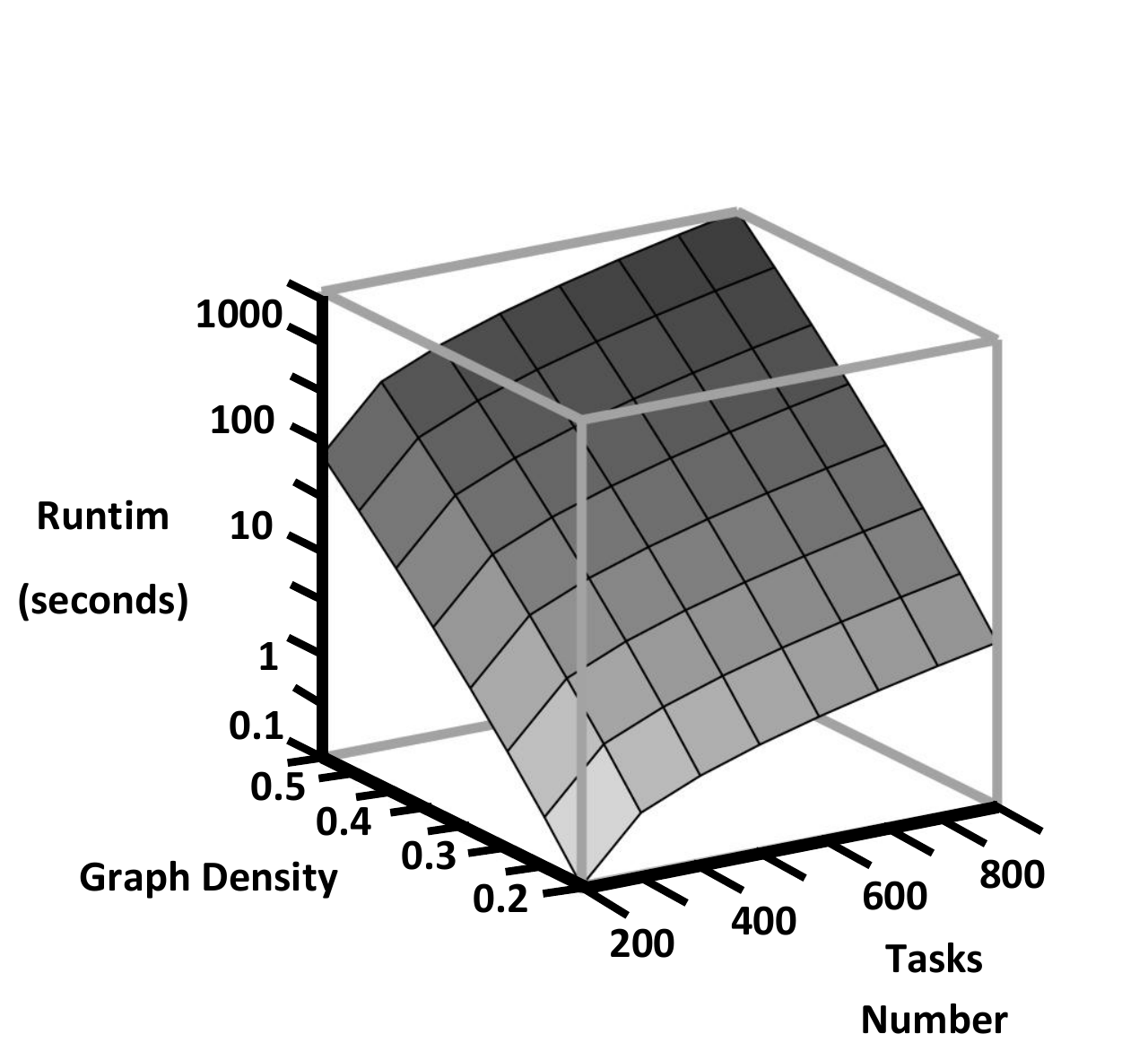} 
\caption{Proposed approach runtime evolution} 
\label{3dcurve} 
\end{center} 
\end{figure}

The first test concerns time performances of the proposed approach functions of the graph\rq{}s number of tasks and the graph\rq{}s density. The diagram of figure \ref{3dcurve} depicts the evolution of the runtime by a 3d curve. It showed that the increasing density has a  more important impact, than those of the number of graph tasks, on the runtime of the proposed approach. This is mainly explained by the fact that the number of paths increases when the graph has a higher density.  Moreover, the runtime of the proposed approach are very reasonable even when the density is  hight.  Notice that the runtime follows a logarithmic scale and results were collected on a machine with a 3,4 GHz Intel Core i7 processor
and 10GB main memory.

The second test targets the efficiency of the proposed approach in term of schedulability and lower bounds. To do so, we chose to use the constraint programming for resolving the latency constraints scheduling problem and to compare the obtained results to the proposed approach results. 

The constraint programming is a programming language that is oriented to relationships or constraints among entities \cite{Apt:2003:PCP:1237975}.  The most important reason is that constraint programming has a rich modeling language which is very convenient to express the problem. Moreover, the underlying CP solver is relatively robust with respect to the addition of new constraints, and the search can be controlled entirely by the user.

Our problem was solved using ILOG OPL Studio commercial software according to the following CP formulation. The objective is to minimize the scheduling of tasks under $L_1$ by minimizing the start time of $t_b$ and in the same time minimizing the scheduling of tasks under $L_2$ by minimizing the start time of $t_d$  (knowing that latency constraint are imposed on $(t_a,t_b)$ and $(t_c,t_d)$). Hence, the multiple objectives are expressed in a single objective by summing them together and applying weights to each objective to signify its relative importance. It was assessed, first, that the two objectives have the same importance and, consequently, the same weight. But, the runtime of CP approach exploded, even for small graphs. Hence, CP approach minimizes $L_1$ first then $L_2$. Thus, the objective function is:  
\begin{equation}\nonumber
\mathrm{Min} \ \ (x*\mathrm{StartOf}(t_b) + y*\mathrm{StartOf}(t_d))
\end{equation}
Where $(x,y)=(1,0)$ then $(x,y)=(0,1)$.  In addition, variables domains and constraints are given in table \ref{table1} and \ref{table2}. Constraints of table \ref{table2} are provided by ILOG OPL Studio for scheduling modeling \cite{ilog}.  The number of processors is defined by Algorithm \ref{algo1}.

\begin{table}[tp] %
\caption{Definition of Variables and Domains}
\label{table1}\centering %
\begin{tabular}{l|l}
\hline    
Variable  & Domain    \\  \hline
NbTasks& $\mathbb{N}^+$       \\ 
NbProcs  & $\mathbb{N}^+$       \\
duration($t_i$)  & $\mathbb{N}^+$    \\ 
task$(t_i,proc_j)$ &  [StartOf($t_i$), EndOf($t_i$)] $\subset \mathbb{N}^+$  \\  \hline
\end{tabular}
\end{table}

\begin{table}[tp] %
\caption{Definition of Constraints}
\label{table2}\centering %
\begin{tabular}{l|l}
\hline
Constraint  &  Description \\ \hline
$\cdot$ if $(t_i,t_j)\in \mathbb{E}$ &  $\cdot$ $t_j$ is a predecessor of  $t_i$  \\  
then EndOf($t_i$) $\leq$ StartOf($t_j$)  & \\
$\cdot$$\forall t_i \in   \mathbb{V}$, $\forall  proc_j$,  & $\cdot$ each task needs only one \\ 
 alternative(task$(t_i,proc_j)$) & processor to be executed \\
$\cdot$ $ \forall proc_i$,  noOverlap($proc_i$) &   $\cdot$ no overlap on processors \\ \hline 
\end{tabular}
\end{table}

To do so, within the CP approach the objective was to look for the scheduling which minimizes the start dates of $t_b$ and $t_d$ then to
compute the values of $L_{1}^{opt}$ = StartOf$(t_b)$ and $L_{2}^{opt}$=StartOf$(t_d)$. These values are the 
optimal (smallest) values that $L_1$ and $L_2$ can have. Then, they were compared to the values of $L_{1}^{lb}$ and $L_{2}^{lb}$ resulting from the calculation of equations \ref{lb}.

\begin{figure}[h!] 
\begin{center} 
\includegraphics[scale=1]{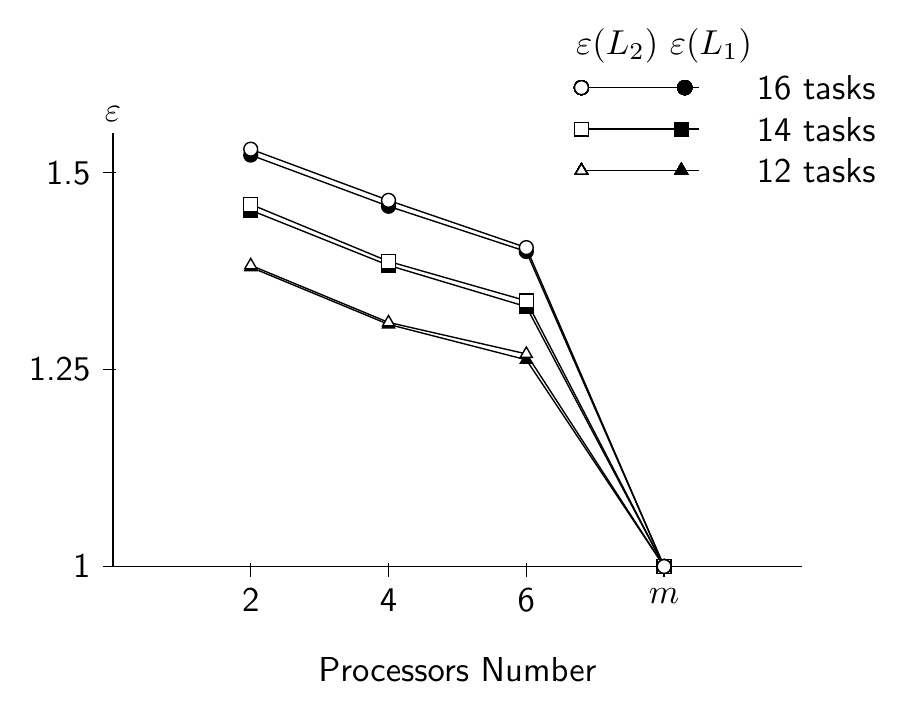} 
\caption{Proposed approach schedulability performances} 
\label{plot} 
\end{center} 
\end{figure}
After that, the value of $\rho$ is computed which the ratio between $L_{i}^{opt}$ and $L_{i}^{lb}$ such that  $\rho (L_i) = \frac{L_{i}^{opt}}{L_{i}^{lb}}$ 
 in order to get an  idea of how far are the proposed approach results from the optimal ones. For each case of the tasks number list [12,14,16] until 20 different graphs were generated and both approaches were applied on them. Notice that the chosen density of all tested graphs was 0.4. At the beginning, the two approaches were executed on a $m$ processors architecture ($m$ is given by Algorithm \ref{algo1}). After that, the number of processor was reduced and fixed from the list [4,3,2], and only the optimal approach was executed. Notice that the proposed approach cannot be execute since it fixes itself  the number of processors. Results are illustrated by diagrams on figure \ref{plot} ($\rho (L_1)$ is marked in black and $\rho (L_2)$ is in white). 
 
As expected, $\rho$ is equal to 1 when the number of processors is equal to $m$ meaning that our approach as the optimal approach return the optimal latency values in the case of $m$ processors. After that, once the number of processors was reduced, $\rho$ values increase meaning that the values returned by CP is larger than $L_{i}^{lb}$ and $L_{i}^{lb}$  which confirm their positions of lower bounds. Notice that the values of $\rho$ increase, also, from the first set of tests (12 tasks) to the second set of tests (14 tasks) then increase again in the third set of tests (16 tasks). This is explained by the fact that when the number of tasks increase, the number of paths follows and it leads to increase the value of the optimal number of processors $m$. 
 
In addition, it emerges that the proposed approach provides an interesting results considering that, among the three sets of tests,  optimal approach returned results varying from ~1.25 and ~1.5 times the proposed approach results. This means that, on all performed experiments, the proposed approach gives a value to $L_i$ which is, at worst, around 1.5 times smaller than the one given by the CP approach. Hence, the proposed lower bounds can be considered as efficient seeing the difference between runtimes of the two approaches.  The light difference between $\rho(L_1)$ and $\rho(L_2)$ is explained by the fact that in CP approach priority is given to the minimization of (StartOf$(t_b)$) at the cost of minimization of  (StartOf$(t_d)$). As with any other optimal method,  runtime of CP approach explodes exponentially  as soon as the number of tasks becomes more important which prevented us to consider more than 16 tasks graphs.

\section{Conclusion} The paper presents a theoretical study of the real-time non preemptive multiprocessor scheduling with precedence and several latency constraints. After assessing the NP-hardness of this problem, an algorithm is proposed for allocating application graph tasks to a number of processors allowing the optimal task parallelism exploitation. 
The schedulability study, proposed here, introduces a first condition in the case of one latency constraint. Then, after giving the different possible configurations in the case of several latency constraints, it introduces a second condition to check the schedulability of latency constraints in the hardest configuration in term of complexity. Finally, from the proposed conditions a practical lower bounds were deduced.       

The first phase of tests demonstrates that the proposed approach has a very competitive runtime. In addition, the second phase concerned a comparison with an optimal approach which is the Constraint Programming approach. These tests showed that the proposed approach provides an interesting results in term of schedulability and lower bounds.     

The performed study assumes that the number of processors is at least equal to the number of paths selected by the allocation algorithm. Hence, it is plan to explore the possibilities
of including the number of processors in the schedulability condition as a fixed parameter.

\bibliographystyle{unsrt} 
\bibliography{ref,bibSyndex}

\begin{thebibliography}{10}

\bibitem{Amdahl67}
G.~M. Amdahl.
\newblock Validity of the single-processor approach to achieving large scale
  computing capabilities.
\newblock In {\em AFIPS Conference Proceedings}, volume~30, pages 483--485.
  AFIPS Press, 1967.

\bibitem{Shi1996}
Yuan Shi.
\newblock Reevaluating amdahl's law and gustafson's law.
\newblock Technical report, Temple University, Philadelphia, PA 19122, October
  1996.

\bibitem{jeffaynonpreemptive}
K.~Jeffay, D.~F. Stanat, and C.~U. Martel.
\newblock On non-preemptive scheduling of periodic and sporadic tasks.
\newblock In {\em Proceedings of the 12 th IEEE Symposium on Real-Time
  Systems}, pages 129--139, December 1991.

\bibitem{SERTS}
F.~Balarin, L.~Lavagno, P.~Murthy, and A.~Sangiovanni-vincentelli.
\newblock Scheduling for embedded real-time systems.
\newblock {\em IEEE Design and Test of Computers}, 15(1):71--82, 1998.

\bibitem{10.1109/RTCSA.2011.72}
Cong Liu and James~H. Anderson.
\newblock Supporting graph-based real-time applications in distributed systems.
\newblock {\em Real-Time Computing Systems and Applications, International
  Workshop on}, 1:143--152, 2011.

\bibitem{DBLP:conf/rtss/BaruahG08}
Sanjoy~K. Baruah and Joel Goossens.
\newblock The edf scheduling of sporadic task systems on uniform
  multiprocessors.
\newblock In {\em IEEE Real-Time Systems Symposium}, pages 367--374, 2008.

\bibitem{DBLP:conf/ecrts/BaruahL04}
Sanjoy~K. Baruah and Giuseppe Lipari.
\newblock Executing aperiodic jobs in a multiprocessor constant-bandwidth
  server implementation.
\newblock In {\em ECRTS}, pages 109--116, 2004.

\bibitem{DBLP:conf/rtcsa/HuangCT11}
Kai Huang, Jian-Jia Chen, and Lothar Thiele.
\newblock Energy-efficient scheduling algorithms for periodic power management
  for real-time event streams.
\newblock In {\em RTCSA (1)}, pages 83--92, 2011.

\bibitem{journals/rts/AbeniCLMP05}
Luca Abeni, Tommaso Cucinotta, Giuseppe Lipari, Luca Marzario, and Luigi
  Palopoli.
\newblock Qos management through adaptive reservations.
\newblock {\em Real-Time Systems}, 29(2-3):131--155, 2005.

\bibitem{DBLP:journals/tii/ButtazzoBW11}
Giorgio~C. Buttazzo, Enrico Bini, and Yifan Wu.
\newblock Partitioning real-time applications over multicore reservations.
\newblock {\em IEEE Trans. Industrial Informatics}, 7(2):302--315, 2011.

\bibitem{papierok}
O.~Kermia.
\newblock Optimizing distributed real-time embedded system handling dependence
  and several strict periodicity constraints.
\newblock {\em Advances in Operations Research}, page 10.1155/2011/561794,
  2011.

\bibitem{Goddard98onthe}
S.~M. Goddard and Jr.
\newblock On the management of latency in the synthesis of real-time signal
  processing systems from processing graphs, 1998.

\bibitem{Hsueh_Lin_2000}
Chih-wen Hsueh and Kwei-jay Lin.
\newblock Scheduling real-time systems with end-to-end timing constraints using
  the distributed pinwheel model.
\newblock {\em IEEE Transactions on Computers}, 49(1):51--66, 2000.

\bibitem{aor07}
L.~Cucu, N.~Pernet, and Y.~Sorel.
\newblock Periodic real-time scheduling: from deadline-based model to
  latency-based model.
\newblock {\em Annals of Operations Research}, 2007.

\bibitem{Chao94minimizingredundant}
H-Yi Chao and M~P. Harper.
\newblock Minimizing redundant dependencies and interprocessor
  synchronizations.
\newblock {\em International Journal of Parallel Programming}, 23:245--262,
  1994.

\bibitem{Cucinotta11-SOMRES}
Tommaso Cucinotta.
\newblock Optimum scalability point for parallelisable real-time components.
\newblock In {\em Proceedings of the International Workshop on Synthesis and
  Optimization Methods for Real-time and Embedded Systems (SOMRES 2011)},
  Vienna, Austria, November 2011.

\bibitem{Book:IntroductionToParallelComputing}
Lawrence Livermore National~Laboratory Blaise~Barney.
\newblock Introduction to parallel computing.
\newblock Web, 2010.

\bibitem{grah79}
R.~L. Graham, E.~L. Lawler, J.~K. Lenstra, and A.~H. G.~Ronnooy Kan.
\newblock Optimization and approximation in deterministic sequencing and
  scheduling: a survey.
\newblock In {\em Annals of Discrete Mathematics}, 1979.

\bibitem{VisSchKonBor10}
S.~V.~N. Vishwanathan, N.~Schraudolph, R.~Kondor, and K.~Borgwardt.
\newblock Graph kernels.
\newblock {\em Journal of Machine Learning Research}, 11:1201--1242, 2010.

\bibitem{Tutzauer2007}
F~Tutzauer.
\newblock Entropy as a measure of centrality in networks characterized by
  path-transfer flow.
\newblock {\em Social Networks}, 29(2), 2007.

\bibitem{YuchunMa:2007}
Yuchun Ma, Zhuoyuan Li, Jason Cong, Xianlong Hong, G.~Reinman, Sheqin Dong, and
  Qiang Zhou.
\newblock Micro-architecture pipelining optimization with throughput-aware
  floorplanning.
\newblock In {\em Proceedings of the 2007 Asia and South Pacific Design
  Automation Conference}, 2007.

\bibitem{cucu04}
L.~Cucu and Y.~Sorel.
\newblock Non-preemptive scheduling algorithms and schedulability conditions
  for real-time systems with precedence and latency constraints.
\newblock (RR-5403):33, 2004.

\bibitem{GlaBer:2010}
Christian Glasser, Christian Reitwiessner, Heinz Schmitz, and Maximilian Witek.
\newblock Approximability and hardness in multi-objective optimization.
\newblock In {\em Proceedings of the Programs, proofs, process and 6th
  international conference on Computability in Europe}, CiE'10, 2010.

\bibitem{Apt:2003:PCP:1237975}
Krzysztof Apt.
\newblock Principles of constraint programming.
\newblock 2003.

\bibitem{ilog}
Philippe Laborie.
\newblock Ibm ilog cp optimizer for detailed scheduling illustrated on three
  problems.
\newblock In {\em Integration of AI and OR Techniques in Constraint Programming
  for Combinatorial Optimization Problems}, Lecture Notes in Computer Science.
  2009.

\end{thebibliography}

\end{document}